\documentclass[10pt,conference]{IEEEtran}

\usepackage[utf8]{inputenc}
\usepackage{cite}
\usepackage{amsmath,amssymb,amsfonts}
\usepackage{algorithmic}
\usepackage{graphicx}
\usepackage[export]{adjustbox}
\usepackage{textcomp}
\usepackage{xcolor}
\usepackage{hyperref}
\usepackage{dblfloatfix}

\usepackage{booktabs}
\usepackage{multirow}

\def\BibTeX{{\rm B\kern-.05em{\sc i\kern-.025em b}\kern-.08em
    T\kern-.1667em\lower.7ex\hbox{E}\kern-.125emX}}

\IEEEoverridecommandlockouts\IEEEpubid{\makebox[\columnwidth]{\hfill} \hspace{\columnsep}\makebox[\columnwidth]{ }}

\begin{document}

\title{Empirical Time Complexity of \\ Generic Dijkstra Algorithm}

\author{
\IEEEauthorblockN{Piotr Jurkiewicz\IEEEauthorrefmark{1}, Edyta Biernacka, Jerzy Domżał and Robert Wójcik}
\IEEEauthorblockA{\textit{Department of Telecommunications} \\
\textit{AGH University of Science and Technology}\\
Kraków, Poland \\
\IEEEauthorrefmark{1}piotr.jurkiewicz@agh.edu.pl}
}

\maketitle

\begin{abstract}

Generic Dijkstra is a novel algorithm for finding the optimal shortest path in both wavelength-division multiplexed networks (WDM) and elastic optical networks (EON), claimed to outperform known algorithms considerably. Because of its novelty, it has not been independently implemented and verified. Its time complexity also remains unknown. In this paper, we perform run-time analysis and show that Generic Dijkstra running time grows quadratically with the number of graph vertices and logarithmically with the number of edge units. We also discover that the running time of the Generic Dijkstra algorithm in the function of network utilization is not monotonic, as peak running time is at approximately 0.25 network utilization. Additionally, we provide an independent open source implementation of Generic Dijkstra in the Python language. We confirm the correctness of the algorithm and its superior performance. In comparison to the Filtered Graphs algorithm, Generic Dijkstra is approximately 2.3 times faster in networks with 25 to 500 nodes, and in 90\% of calls its computation takes less time.

\end{abstract}


\section{Introduction}

The shortest-path Dijkstra algorithm is known as an optimal and efficient solution to find a shortest path in a graph. In optical networks, finding a path to accommodate a given traffic demand is more challenging due to wavelength/spectrum assignment constrains.

In wavelength-division multiplexed (WDM) networks this is known as the routing and wavelength assignment (RWA) problem. In elastic optical networks (EON), this problem evolves into the routing and spectrum assignment (RSA) or the routing, modulation and spectrum assignment (RMSA) problem.

The mentioned problems can be interpreted in two ways:
\begin{itemize}
\item to ensure the optimal global network performance which can be expressed as the minimum bandwidth-blocking probability for a group of demands (equivalent to the \emph{multi-commodity flow problem} in graph theory),
\item to ensure optimality of a single connection, which means finding the shortest path capable of supporting a given demand (equivalent to the \emph{shortest path problem}).
\end{itemize}

While the first version (called \emph{static} or \emph{offline}) is NP-complete, the second one (called \emph{dynamic} or \emph{online}) can be solved tractably. A well known approach is to use the so-called \emph{Filtered Graphs} algorithm, which solves the dynamic RWA, RSA or RMSA problems by finding shortest paths (using the Dijkstra algorithm) in a number of graphs containing only subset of these edges which can support a given slot determined according to the demand and available modulation formats and then selecting the best of them.

Recently, the \emph{Generic Dijkstra} algorithm has been proposed in \cite{Szczesniak:19} as an alternative. According to its authors, it finds optimal solutions for RWA, RSA and RMSA problems, and at the same time is considerably faster than the Filtered Graphs algorithm.

The goal of the research presented in this paper is as follows:

\begin{itemize}

\item to independently implement the Generic Dijkstra algorithm, basing on the original description, and verify its accuracy and performance,

\item to verify authors' claims regarding its superior speed compared to the Filtered Graphs algorithm,

\item and, most importantly, to investigate how its running time depends on input network parameters, as the complexity analysis was not presented by its authors.

\end{itemize}

This paper provides a valuable insight into the performance of the Generic Dijkstra algorithm. We show the simulation results for both approaches (Filtered Graphs and Generic Dijkstra), compare their speed and determine empirical orders of growth of average call time depending on network size, the number of units on each edge and network utilization\footnote{Network utilization is defined as the ratio of the number of units in use in the network to the total number of units on all edges.}. We also provide the implementation and tests in an open source repository \cite{github-generic-dijkstra}.

\section{Related works}\label{related-works}

EON have been introduced as flexible and heterogeneous concept to replace WDM \cite{Gerstel2012}. In EON, to achieve such elastic access, a frequency slot is introduced as a unit of dividing optical spectrum resources (instead of one wavelength). The width of this slot corresponds to the bandwidth of the orthogonal frequency-division multiplexing (OFDM) subcarrier. As a result, optical connections employ different modulation formats and occupy only the required number of of slots.

To set up an optical connection between a pair of nodes in EON \cite{Gerstel2012}, the RSA problem needs to be solved. Such a solution must satisfy the following constraints \cite{Christodoulopoulos2011}:

\begin{itemize}

\item spectrum continuity constraint -- the connection must allocate the same slots along links of an end-to-end path,

\item spectrum contiguousness constraint -- all slots assigned to a connection should be adjacent,

\item non-overlapping spectrum constraint -- at the same time, at most one connection occupies spectrum of links.

\end{itemize}

To satisfy all constraints, RSA methods have been proposed for both static and dynamic scenarios \cite{Velasco2014c}. The classification of RSA methods with theoretical descriptions can be found in \cite{Chatterjee2015a_tut}, whereas numerical results are presented in \cite{Zhu2013} and \cite{Biernacka2017b}. For example, in \cite{Zhu2013} the authors proposed a heuristic method for selecting a path with the lowest link utilization. All of these RSA algorithms are based on heuristic methods. This means that, while they are fast, they results are not guaranteed to be optimal. They may return sub-optimal solutions or return no solution at all despite its existence.

Whereas static RSA is NP-complete, dynamic RSA may be solved optimally (but inefficiently) by finding the shortest paths in filtered graphs. The authors of \cite{Szczesniak:19} proposed a novel algorithm, which they called the Generic Dijkstra. In their proposal, the original shortest-path Dijkstra algorithm has been generalized to finding the shortest path in optical networks for a given demand. The generalization resolves the continuity and contiguity constraints for units, while the constriction takes into account constraints of modulation. The detailed operations of the Generic Dijkstra algorithm will not be presented there, as they are presented in the original paper \cite{Szczesniak:19}. The authors conclude that Generic Dijkstra is the first proposal as the optimal and efficient algorithm for the dynamic routing problem.

\section{Implementation}

The original implementation of the Generic Dijkstra algorithm published in \cite{gDijkstraimpl} was coded in C++. We did not use nor consulted that code in our work. Instead, we decided to implement the algorithm from scratch using Python. Our implementation is based solely on the algorithm descriptions presented in the original article. This is important, as it confirms that the description in paper is precise and sufficient to implement the algorithm correctly.

In case of the Filtered Graphs algorithm, we used Dijkstra implementation from the popular \emph{networkx} library \cite{networkx}. We introduced a straightforward optimization based on the idea of inline filtering of edges during Dijkstra algorithm calls \footnote{In the original version of the Filtered Graphs algorithm, the filtering of a network graph (removing edges which cannot support a given continuous set of slots) is performed before each Dijkstra call. All edges in the graph are checked and then the Dijkstra algorithm is called on the subgraph with infeasible edges filtered out. In our implementation, the check whether a particular edge can support given slots is performed inline in the inner loop of the Dijkstra algorithm, when this edge is traversed. Because only a subset of edges are traversed during a typical Dijkstra algorithm call (all edges are traversed only in the worst case, which is the linear graph), the number of checks is always lower in the inline version of the algorithm, which gives performance benefits.}.

We admit that Python is not the most effective language in terms of absolute speed. However, the main goal of our research was to investigate the relative performance of Generic Dijkstra algorithm compared to the Filtered Graphs algorithm and determine the orders of growth. Thus, as both of them were implemented by us in Python, the programming language inefficiency is irrelevant. Moreover, we performed our simulations using two Python 3 runtimes: CPython, the official Python interpreter, and PyPy, which is a JIT runtime. Although the PyPy is much faster, due to just-in-time compilation to machine code, relative performance is similar in both runtimes, which corroborates the assumption that our results can be generalized to other programming languages.

The implementation code and test cases are available at:

\smallskip

\noindent \url{https://github.com/piotrjurkiewicz/generic-dijkstra} \cite{github-generic-dijkstra}

\smallskip


\section{Simulations}

In order to investigate time complexity as a function of the number of nodes, we had to perform simulations on many topologies of different sizes. Because of that, the well-known classical topologies, like NSF network, were not sufficient for us. Instead, we generated 10 different Gabriel graphs for 20 different graph sizes (from 25 to 500 vertices), which gave the total number of 200 different topologies. Gabriel graphs have been shown to model the properties of the long-haul transport networks very well \cite{6798402}. These topologies are available in the code repository \cite{github-generic-dijkstra}. The number of graph edges was not considered as an input parameter, because in Gabriel graphs it depends on the location of vertices and cannot be controlled directly.

For each topology, we performed simulations assuming 10 different number of units available on edges (from 100 to 1000 units on each edge). All simulations were repeated for 2 different mean numbers of demanded units (10\% and 5\% of edge available units) and with 10 different seeds controlling demand generator. All the above resulted in a total of 40000 simulations for each algorithm-runtime combination.

In each simulation, the Filtered Graphs or the Generic Dijkstra algorithm was called in loop until the network utilization reached 0.6. At that moment simulation was stopped.

We assumed 4 modulation levels ($M=4$), similarly as in most EON papers. The reach of the least spectrally efficient modulation was set to 1.5 of the longest shortest path in each particular topology. The parameters of simulation server are shown in Table \ref{tab-meta}.

\begin{table}[!h]
\caption{Simulation server parameters}
\begin{center}
\begin{tabular}{@{}lr@{}}
\toprule
\textbf{Distribution} & Debian Buster \\
\textbf{Kernel version} & 4.19.98-1 \\
\textbf{CPython version} & Python 3.7.3 \\
\textbf{PyPy version} & PyPy 7.0.0 with GCC 8.2.0 \\
\textbf{CPU} & Intel Xeon E5-2690 \\
\textbf{CPU Frequency} & 3300 MHz \\
\bottomrule
\end{tabular}
\label{tab-meta}
\end{center}
\end{table}

\section{Results}

\begin{figure*}[!t]
\centering
\includegraphics[width=\textwidth]{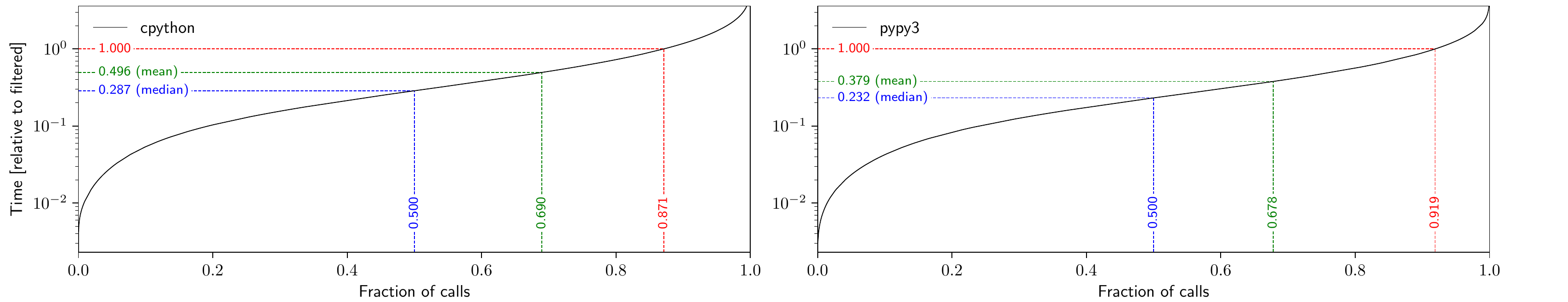}
\caption{Performance of Generic Dijkstra relative to Filtered Graphs algorithm. CPython on left, PyPy on right.}
\label{relative_to_filtered_cdf}
\end{figure*}

The first objective of our research was to validate the correctness and optimality of Generic Dijkstra algorithm. We checked that for all 51782749 calls it yields exactly the same results (the same set of paths) as the Filtered Graphs algorithm. This confirms that the Generic Dijkstra algorithm provides optimal solutions and can be implemented correctly based on its description in the original article.

The second objective was to verify how fast is Generic Dijkstra compared to Filtered Graphs. In Figure \ref{relative_to_filtered_cdf} we present the cumulative distribution of time taken by Generic Dijkstra calls compared to Filtered Graphs calls. We compare calls at the same moment of the simulation (i.e. operating on the same state of the network) to each other.

It can be seen that Generic Dijkstra is on average 2.02 (running on CPython) or 2.64 (running on PyPy) times faster than the Filtered Graphs algorithm. For 50\% of calls it is at least 3.48 (CPython) or 4.31 (PyPy) times faster. For 87.1\% of calls on CPython and 91.9\% calls on PyPy Generic Dijkstra provides the solution faster than the Filtered Graphs algorithm.

\begin{figure*}[!h]
\centering
\includegraphics[width=\textwidth]{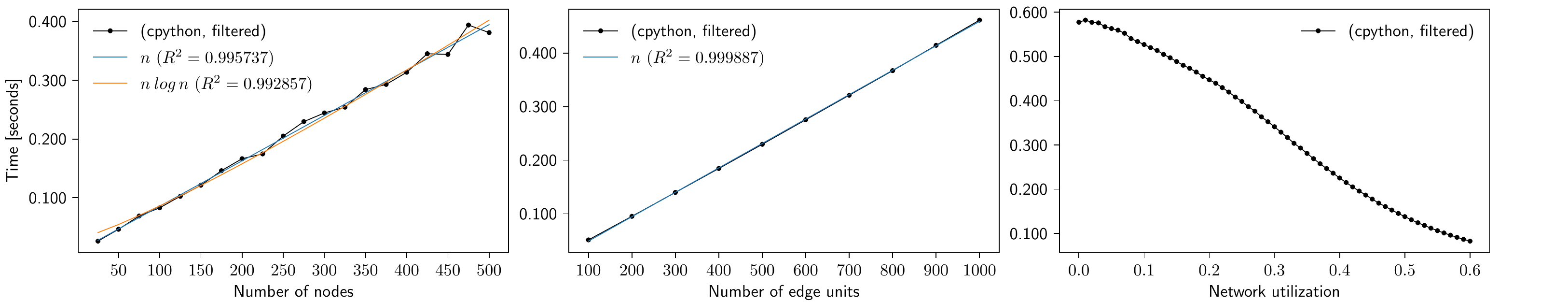}
\caption{Performance of Filtered Graphs algorithm. CPython.}
\label{cpython_filtered_groups}
\end{figure*}

\begin{figure*}[!h]
\centering
\includegraphics[width=\textwidth]{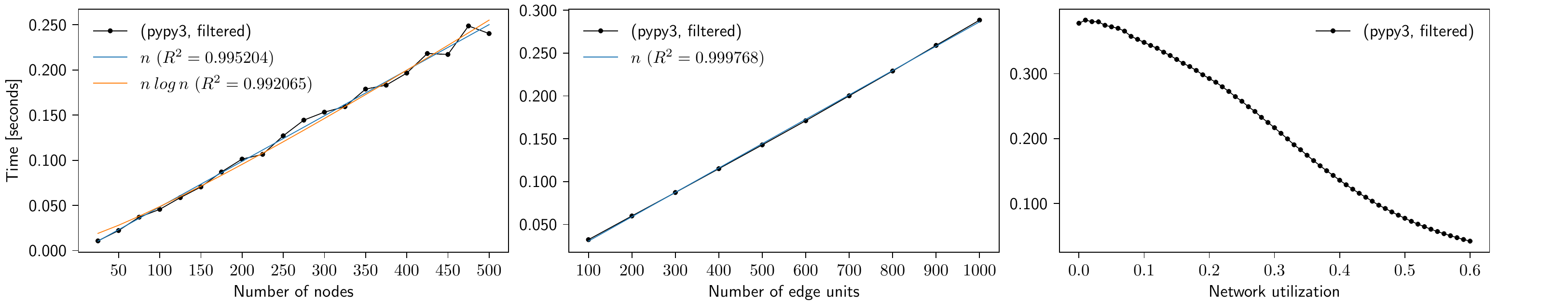}
\caption{Performance of Filtered Graphs algorithm. PyPy.}
\label{pypy3_filtered_groups}
\end{figure*}

\begin{figure*}[!h]
\centering
\includegraphics[width=\textwidth]{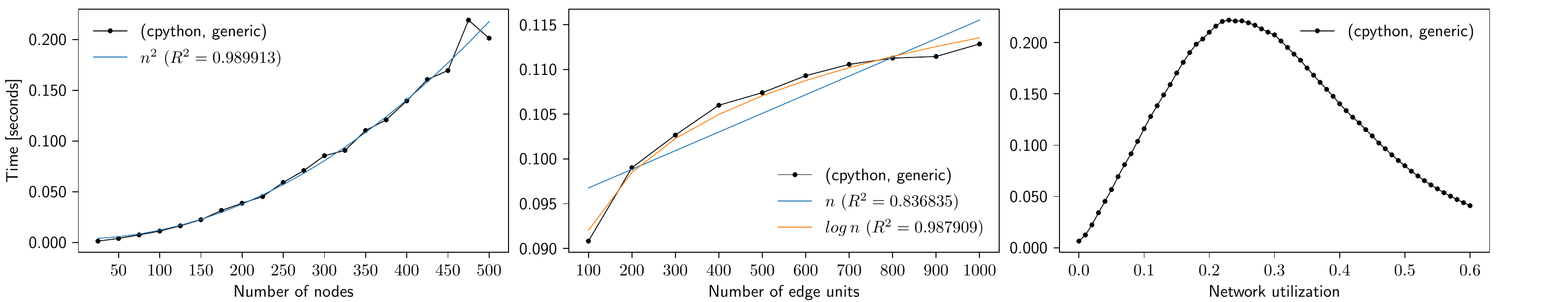}
\caption{Performance of Generic Dijkstra algorithm. CPython.}
\label{cpython_generic_groups}
\end{figure*}

\begin{figure*}[!h]
\centering
\includegraphics[width=\textwidth]{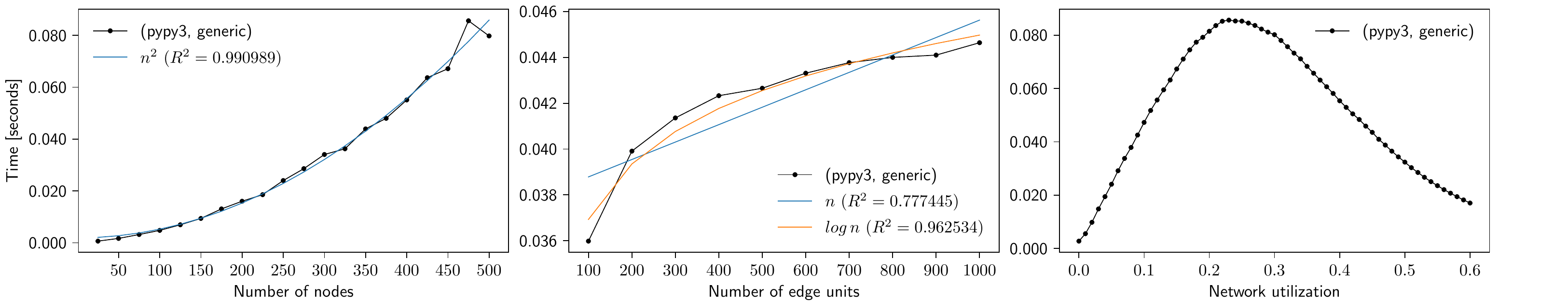}
\caption{Performance of Generic Dijkstra algorithm. PyPy.}
\label{pypy3_generic_groups}
\end{figure*}

In Figures \ref{cpython_filtered_groups}, \ref{pypy3_filtered_groups}, \ref{cpython_generic_groups} and \ref{pypy3_generic_groups}, we present the average call time depending on the network size, the number of units and network utilization for the both interpreters interpreter. Additionally, we try to fit curves of different mathematical functions to determine the orders of growth.

In case of the Filtered Graphs algorithm, its average time complexity can be determined analytically and equals $O(S \: V \: log\,V)$, where $S$ is the number of edge units and $V$ is the number of vertices in the graph. Figures \ref{cpython_filtered_groups} and \ref{pypy3_filtered_groups} shows empirically determined time complexities, which are in line with analytical values. With increasing network utilization, running time of the algorithm decreases linearly. This can be attributed to the fact, that when network utilization is higher, more Dijkstra calls return early when edges close to the source cannot support the selected set of slots.

In case of the Generic Dijkstra algorithm, we were unable to determine the average time complexity analytically due to its complex dependencies on several graph features. The pessimistic complexity analysis can be performed, however, it does not give a insight into the performance of algorithm in realistic networks. Figures \ref{cpython_generic_groups} and \ref{pypy3_generic_groups} show empirically determined time complexities. The running time exhibits a quadratic growth rate for network size ($t \sim V^{2}$). On the other hand, with the increasing number edge units, the running time of the algorithm grows logarithmically ($t \sim log\,S$). Interestingly, the running time in function of network utilization is not monotonic. It exhibits a peak around network utilization equal to 0.25 and then decreases.

\begin{figure*}[!h]
\centering
\includegraphics[width=\textwidth]{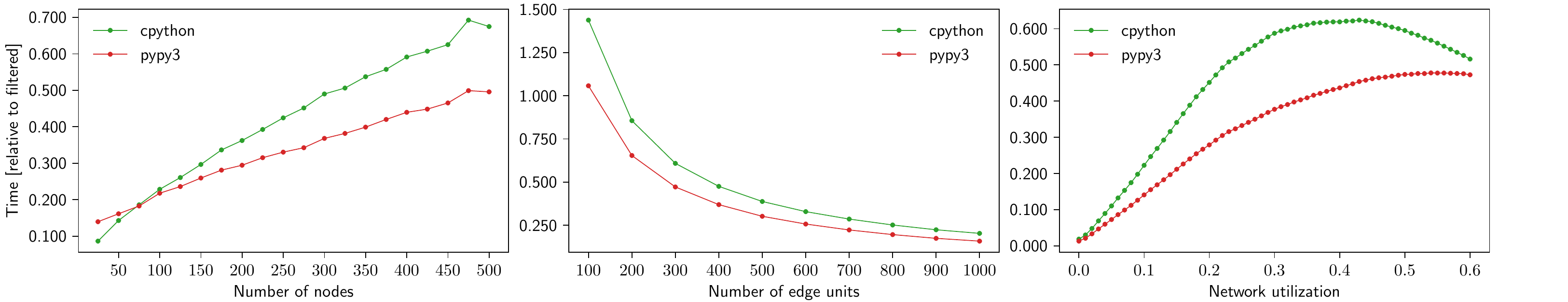}
\caption{Performance of Generic Dijkstra relative to Filtered Graphs algorithm. CPython and PyPy.}
\label{relative_to_filtered_groups}
\end{figure*}

Finally, Figure \ref{relative_to_filtered_groups} presents averaged ratio of Generic Dijkstra call time to Filtered Graphs call time. Because Generic Dijkstra exhibits higher growth rate for network size ($n^{2}$ vs. $n\:log\,n$), its call time approaches Filtered Graphs time as the number of nodes in network increases. However, due to lower constant factors in the algorithm, for network with 500 nodes it still consumes 30-50\% less computation time than Filtered Graphs. Network sizes with more than 500 nodes are too big to be currently considered in EONs. On the other hand, with the increasing number of edge units and network utilization, Generic Dijkstra becomes faster than Filtered Graphs.

\section{Conclusions}

The contribution of this paper is threefold. Firstly, we have independently implemented the Generic Dijkstra algorithm in Python and verified its correctness. We share this implementation as an open source. This confirms that the Generic Dijkstra algorithm indeed works and yields expected results.

Secondly, we compared its performance to the Python implementation the of Filtered Graphs algorithm. We confirmed that for approximately 90\% of calls Generic Dijkstra is faster than Filtered Graphs. On average, Generic Dijkstra is 2.3 times faster than Filtered Graphs. We also presented a CDF graph of relative call time, which gives a good insight into the running time of both algorithms. This analysis was not provided by the author of Generic Dijkstra in his original paper.

Finally, we carried out a run-time analysis and determined empirical orders of growth of the Generic Dijkstra algorithm. They are $t \sim V^{2}$ for the number of graph vertices and $t \sim log\,S$ for the number of edge units. We also discovered that its running time in function of network utilization is not monotonic, with a peak running time at approximately 0.25 network utilization. This is a novel contribution, as no one has yet presented a time complexity analysis of the Generic Dijkstra algorithm.

\section*{Acknowledgment}
\noindent
The research was carried out with the support of the project "Intelligent management of traffic in multi-layer Software-Defined Networks" founded by the Polish National Science Centre under project no. 2017/25/B/ST6/02186.

\bibliographystyle{IEEEtran}

\newpage
\bibliography{./bib}

\end{document}